\documentclass[prl,twocolumn,superscriptaddress]{revtex4-2}   
\usepackage{graphicx,amssymb}
\usepackage{color,ulem}
\usepackage{bm}   

\usepackage{bm}% bold math
\usepackage{amssymb,amsmath}
\usepackage{amsthm}

\usepackage{latexsym}
\usepackage{graphicx,amssymb}
\usepackage{color,ulem}
\usepackage{float}
\usepackage{siunitx}
\usepackage{mathrsfs} 
\usepackage{soul}
\usepackage{graphicx}% Include figure files
\usepackage{dcolumn}% Align table columns on decimal point
\usepackage{bm}% bold math
\usepackage[utf8]{inputenc}
\usepackage{amsfonts,amssymb,amsmath}
\usepackage{epsfig}

\begin{document}
	% \preprint{AIP/123-QED}
	
	\title{%A linear Josephson junction array is not a resistor OR Quantum environments in superconducting circuits}
	Heat bath in a quantum circuit}
	% \thanks{A footnote to the article title}
	
	\author{Jukka P. Pekola}
	\affiliation{Pico group, QTF Centre of Excellence, Department of Applied Physics,
		Aalto University, P.O. Box 15100, FI-00076 Aalto, Finland}
	
	\author{Bayan Karimi}
	\affiliation{Pico group, QTF Centre of Excellence, Department of Applied Physics,
		Aalto University, P.O. Box 15100, FI-00076 Aalto, Finland}
	\affiliation{QTF Centre of Excellence, Department of Physics, Faculty of Science, University of Helsinki, FI-00014 Helsinki, Finland}

	\date{\today}
	
	\begin{abstract}
		We discuss the concept and realization of a heat bath in solid state quantum systems. First we demonstrate that, unlike a true resistor, a finite one-dimensional Josephson junction array or analogously a transmission line with non-vanishing frequency spacing does not strictly qualify as a Caldeira-Leggett type dissipative environment. We then consider a set of quantum two-level systems as a bath, which can be realized as a collection of qubits. We demonstrate that only a dense and wide distribution of energies of the two-level systems can secure long Poincare recurrence times characteristic of a proper heat bath. An alternative for this bath is a collection of harmonic oscillators, for instance in form of superconducting resonators.
	\end{abstract}

	\maketitle
The question of thermalization in closed quantum systems and the nature of thermal reservoirs are topics of considerable interest \cite{mori18,huse15,zwanzig,reimann16,rigol16,popescu06,rigol07}. However, experimental realizations, in particular in solid-state domain are largely missing \cite{reimann16,chen21}. In this paper we compare different types of reservoirs that can be realized in the context of superconducting quantum circuits. An ideal heat bath is a resistor \cite{kuzmin91,yagi97,lotkhov03,jp15,cattaneo21,astafiev22,subero22}, which can be realized in a straightforward way. But mainly because of the compatibility of the fabrication processes, the circuit QED community typically prefers to mimic resistors or simply to produce high-impedance environments by arrays of Josephson junctions or superconducting cavities \cite{corlevi06,jones13,masluk12,pop14,rastelli18,kuzmin19,leger19,gasparinetti20,kuzmin23}. The advantages of a physical resistor in form of metal film is that it has a truly gapless and smooth absorption spectrum, and on the practical side its temperature can be probed by a standard thermometer \cite{rmp06}. A one-dimensional Josephson junction array, on the contrary, although acting as a high impedance environment \cite{wallraff17,arriola18}, presents well-defined resonances in its absorption spectrum up to the plasma frequency and purely capacitive behavior above it, and cannot thus be considered rigorously as a resistor. Experiments on multimode cavities support our conclusion as they exhibit periodic recoveries of the qubit coupled to them \cite{cleland19}. In order to realize a Caldeira-Leggett type true reservoir \cite{caldeira83,leggett87} out of superconducting elements, we propose an ensemble of qubits or $LC$-resonators with a distribution of energies among them.
%\begin{figure}
%	\centering	\includegraphics [width=0.8\columnwidth] {Fig-1-v1.pdf}
%	\caption{.
%		\label{fig1}}
%\end{figure}

\begin{figure*}
	\centering
	\includegraphics [width=2\columnwidth] {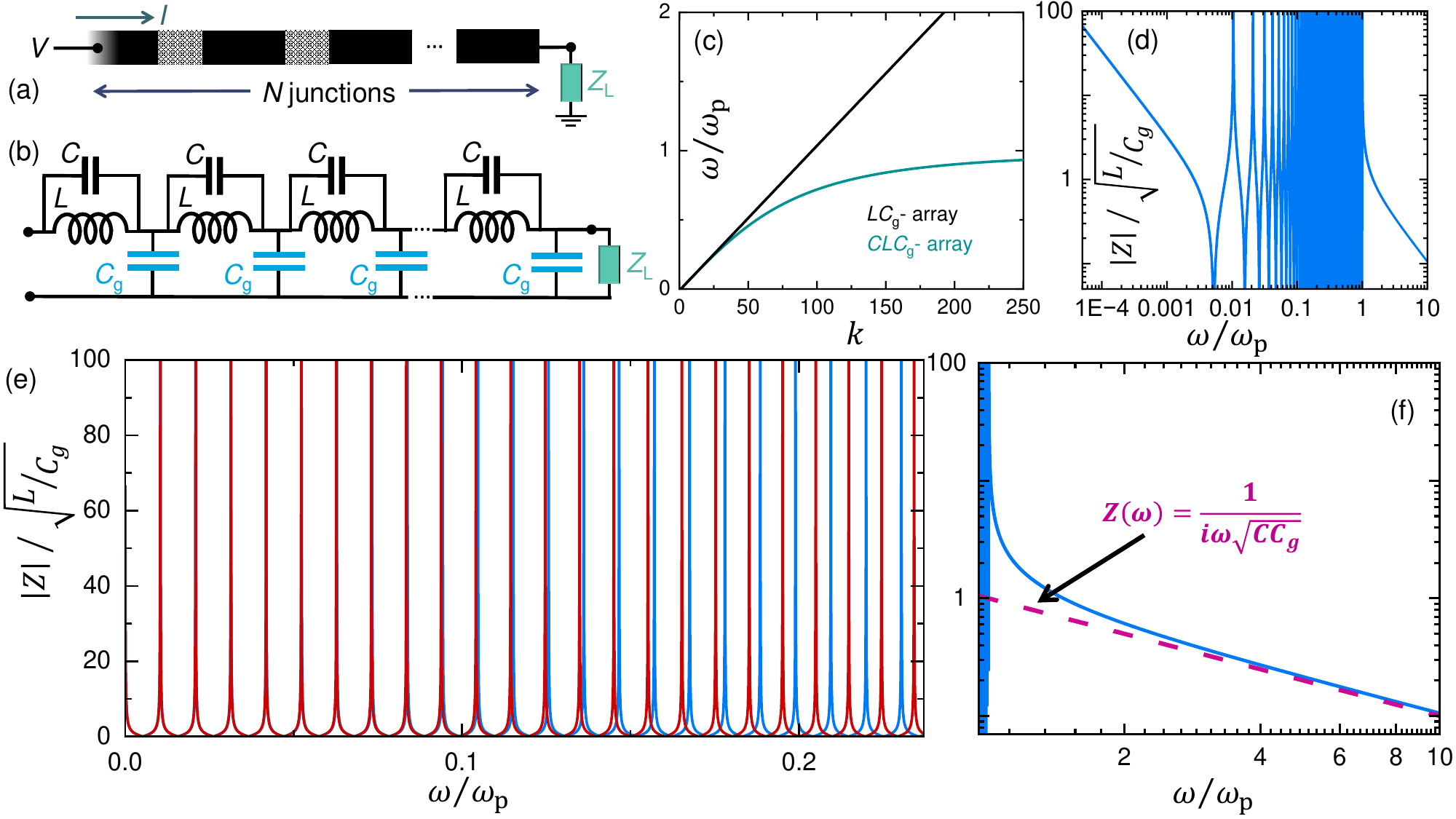}
	\caption{Basic properties of a one-dimensional Josephson junction array. (a) An array with $N$ junctions, terminated by impedance $Z_L$. Current is $I$ and voltage $V$. Junctions can be replaced by superconducting interference devices (SQUIDs) acting as tunable junctions. (b) Equivalent circuit for a uniform array with junctions linearized as inductors $L$. Junction capacitance is $C$ and the stray "ground" capacitance of each island is $C_g$. (c) Dispersion relation for modes in the array for two cases, $C=0$ (black line, $LC_g$) and $C=100C_g$ (green line, $CLC_g$) for an array with $N=3000$. Here we assume an open ended array ($Z_L=\infty$). The (angular) frequencies are scaled by the plasma frequency $\omega_p=1/\sqrt{LC}$ of each junction. (d) Modulus of the impedance of the $CLC_g$ array as a function of frequency, and (e) a zoom out of it for lower frequencies (red line), together with that of the linear $LC_g$ array as well (blue line). (f) At frequencies $\omega \gg \omega_p$, the $CLC_g$ array behaves as a capacitor with effective capacitance $\sqrt{CC_g}$.
		\label{fig1}}
\end{figure*}

We start by an elementary classical analysis of a one-dimensional Josephson junction array (see Fig. \ref{fig1} (a)), which in a linearized form can be presented by a chain of parallel $LC$ elements for the junctions, and a ground capacitance $C_g$ between two of them as in Fig. \ref{fig1} (b). Assuming a long array, we can write for voltage $V(k)$ on island $k$ and current $I(k)$ through the corresponding junction
\begin{eqnarray} \label{eq1}
&& \partial V(k)/\partial k + Z_{LC} I(k)=0 \nonumber \\ && i\omega C_gV(k) + \partial I(k)/\partial k =0.
\end{eqnarray}   
Here $Z_{LC}=-i Z_\infty /(\omega/\omega_p -\omega_p/\omega)$ with $Z_\infty=\sqrt{L/C}$, $\omega$ is the angular frequency of driving, $\omega_p=1/\sqrt{LC}$ is the plasma frequency of the junction, and $I(k)$ is the current through the $k$:th junction. One can solve these equations with different terminations of the array. One finds the dispersion relation of angular frequencies $\omega_n$ for infinite impedance at 
\begin{equation} \label{eq2}
	\omega_n=\omega_{n,0}/\sqrt{1+(\omega_{n,0}/\omega_p)^2},
\end{equation}
where, for an array of $N$ junctions $\omega_{n,0}=(n-1/2)\pi/(N \sqrt{LC_g})$ for a shorted termination and $\omega_{n,0}=n\pi/(N \sqrt{LC_g})$ for an open line \cite{note}. This is the functional dependence of the dispersion relation used in fitting the data, e.g., in Refs. \cite{leger19,kuzmin23}, and it is depicted in Fig. \ref{fig1} (c) for two different values of $C/C_g$, one for pure $LC$ transmission line $C/C_g=0$, and the other for $C/C_g=100$. Figure \ref{fig1} (d)-(f) shows the modulus of the frequency dependent impedance of an array calculated numerically for $C/C_g=100$. We conclude that such an array can hardly be considered to be a resistor. Resonant absorption at frequencies corresponding to Eq. \eqref{eq2} is presented in experiments as well \cite{leger19,kuzmin23}. At frequencies above $\omega_p$ there are no more modes and the impedance is purely capacitive with impedance $Z(\omega)=(i\omega\sqrt{CC_g})^{-1}$ asymptotically at high frequencies (see Fig. \ref{fig1} (f)). 

We next analyze the energy exchange between the system (here a qubit) and a reservoir to assess whether the latter qualifies as a thermal bath. In general, an ideal array presents a reactive element that cannot dissipate the energy. Such a conclusion can be drawn for instance by analyzing the population of a qubit coupled to the array.
To be concrete, we follow the model in Refs. \cite{cleland19,jb22}, and consider a qubit with energy $\hbar\Omega$ coupled to a bath of $N$ states with energy of the $j$:th one equal to $\hbar\omega_j$. The Hamiltonian of the whole system and bath is given by 
\begin{equation} \label{Hamilton}
	\mathcal{\hat{H}} = \hbar\Omega \hat{a}^\dagger\hat{a}+\sum_{i=1}^N \hbar\omega_i \hat{b}_i^\dagger \hat{b}_i+\sum_{i=1}^{N}\gamma_i(\hat{a}^\dagger\hat{b}_i+\hat{a}\hat{b}_i^\dagger),
\end{equation}
where $\hat{a}=|g\rangle \langle e|$ for the qubit with eigenstates $|g\rangle$ (ground) and $|e\rangle$ (excited) and $\hat{b}_i^\dagger~(\hat{b}_i)$ is the creation (annihilation) operator of the environment modes. The non-interacting Hamiltonian is $\mathcal{\hat{H}}_0=\hbar\Omega \hat{a}^\dagger\hat{a}+\sum_{i=1}^N \hbar\omega_i \hat{b}_i^\dagger \hat{b}_i$. The parameters $\gamma_i$ represent the coupling of the qubit with each state in the environment for the perturbation, which reads in the interaction picture with respect to $\mathcal{\hat{H}}_0$ 
\begin{eqnarray} \label{Vinter}
	\hat{V}_I(t)=\sum_{i=1}^{N}\gamma_i(\hat{a}^\dagger\hat{b}_ie^{i(\Omega-\omega_i)t}+\hat{a}\hat{b}_i^\dagger e^{-i(\Omega-\omega_i)t}).
\end{eqnarray}
The basis that we use is formed of the states of the system and environment as $\{|0\rangle=|1000...0\rangle,~|1\rangle=|0100...0\rangle,...,|i\rangle=|0~0...1^{(\rm i:th)}...0\rangle\}$, where the first entrance refers to the qubit and from the second on to each of the $N$ states in the bath. In what follows we apply this model to both a multimode cavity and spins as environment. 
We choose the initial state of the whole system (qubit and environment) as $|\psi_I(0)\rangle\equiv |0\rangle$. This corresponds to the ground state of the environment (zero temperature, $T=0$) but with the qubit excited. We solve the Schr\"odinger equation $i\hbar \partial_t|\psi_I(t)\rangle=\hat{V}_I(t)|\psi_I(t)\rangle$ in the interaction picture to find the time evolution of the state of the whole system, $|\psi_I(t)\rangle=\sum_{i=0}^{N}\mathcal{C}_i(t)|i\rangle$.
\begin{figure*}
	\centering
	\includegraphics [width=2\columnwidth] {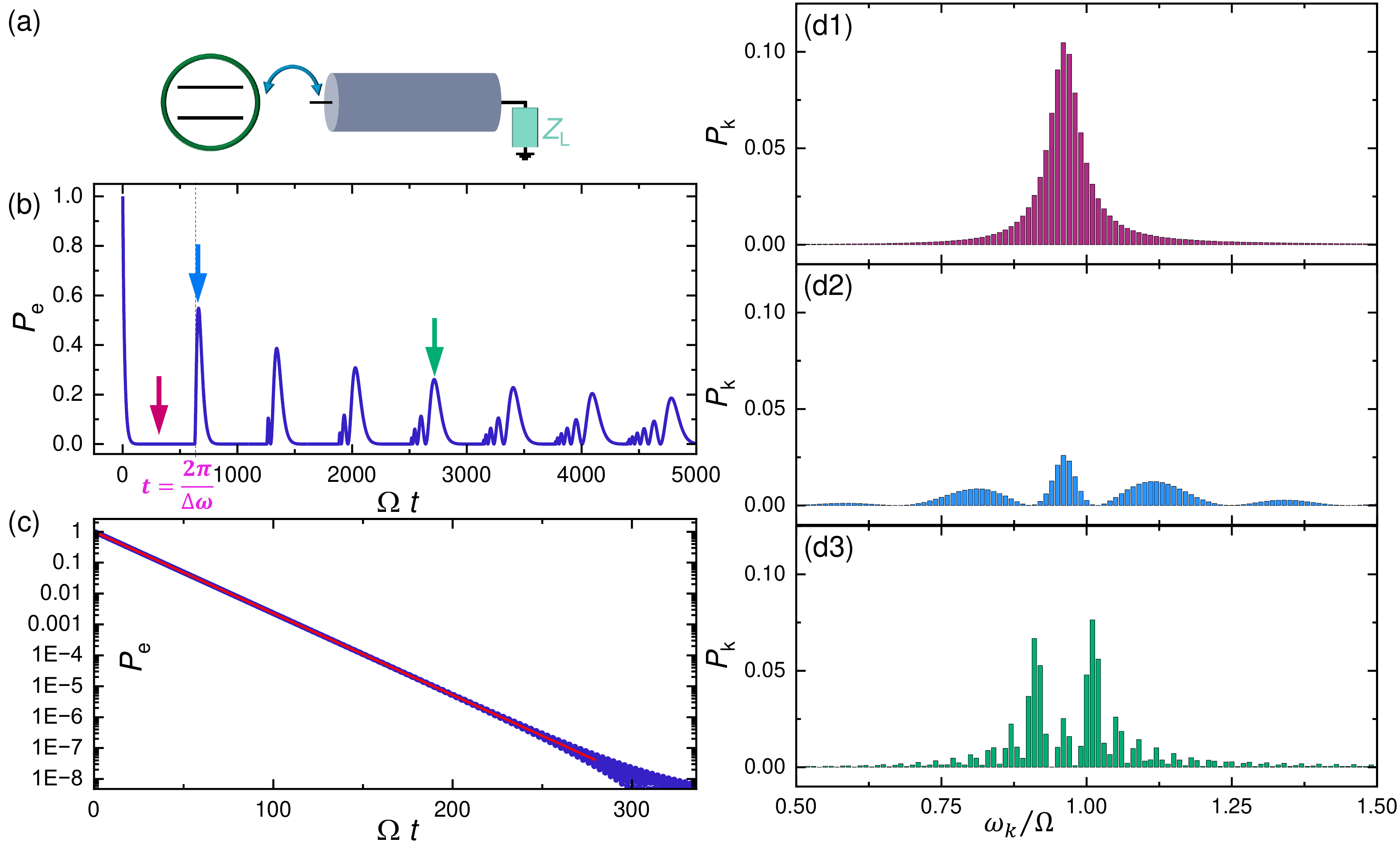}
	\caption{A qubit coupled to a linear Josephson junction array or a transmission line. (a) A schematic presentation of the circuit. (b) Time-dependent population $p_e(t)$ of the qubit after initialization to the excited state. The transmission line is assumed to be initially in the ground state. Coupling parameter between the qubit and the line is $g=0.001$. We have chosen $\Delta \omega = 0.01\Omega$, corresponding to typically either $N=10^4-10^5$ junctions or 1 m long transmission line, close to that in Ref. \cite{cleland19}.  The value of the impedance $Z_L$ has almost no effect on $p_e(t)$. (c) Initially the qubit decays exponentially with decay rate $\Gamma = 2\pi\frac{g^2}{\hbar^2}\frac{\Omega}{\Delta \omega^2}$, until at $t=2\pi/\Delta \omega$ the first revival sets abruptly in. The solid line is an exponential fit in this range. (d) Populations of the states in the multimode resonator at three time instants indicated by arrows in (b). 
		\label{fig2}}
\end{figure*}

Returning first to a Josephson junction array, or a finite transmission line, we may write the (angular) frequencies of the multimode resonator as $\omega_k = k \Delta \omega$ (exactly for an $LC$ transmission line, and approximately for the array well below $\omega_p$, see Eq. \eqref{eq2}), where the spacing $\Delta \omega$ is given by the length of the line or array as discussed above for the latter. Furthermore, we assume the standard coupling as $\gamma_k = g\sqrt{k}$, where $g$ is the coupling constant arising, e.g., from the capacitance between the qubit and the resonator \cite{cleland19}. This model, with the system depicted in Fig. \ref{fig2} (a), demonstrates in the absence of true dissipative elements almost periodic exchange of energy between the qubit and the cavity shown in Fig. \ref{fig2} (b), where the excited state population of the qubit $p_e\equiv |\mathcal C_0|^2$ is depicted against the normalized time $\Omega t$. In this numerical example we chose $\Delta \omega = 0.01\Omega$, and included $N=300$ states in the calculation. This energy spacing mimics approximately the experiment of Ref. \cite{cleland19}.  We can see that the revivals are not full, and the energy of the qubit is distributed over many states with energies in the neighborhood of $\hbar\Omega$. Zooming in to the short time regime as in Fig. \ref{fig2} (c), we observe exponential decay of the population over eight orders of magnitude. A closer analysis of the dynamics yields that indeed the decay in short times is exponential, with a decay rate $\Gamma = 2\pi\frac{g^2}{\hbar^2}\frac{\Omega}{\Delta \omega^2}$, following the numerical result of Fig. \ref{fig2} (c). The other important feature in the dynamics is naturally the periodic recoveries of $p_e(t)$. The first repopulation demonstrates a sharp peak that sets abruptly on at time $t =2\pi /\Delta\omega$. We may associate this with the time of flight of a photon with frequency $\Omega$ through the transmission line and reflected back. In practical circuits this recovery time falls into very short, nanosecond regime, meaning that the transmission line acts as a bath only for times shorter than this. In Ref. \cite{cleland19} similar results as in Fig. \ref{fig2} (b) were obtained using the input-output theory \cite{gardiner85}. The results are robust against different terminations of the line.
   
As is well known, a set of reactive elements can, however, effectively approximate a dissipative element in the spirit of Caldeira and Leggett \cite{caldeira83}. We will next discuss the conditions of forming a heat bath in solid state quantum context without actual dissipative building blocks. In particular we focus on a collection of coupled quantum two-level systems (TLSs), which can in practice be formed of Josephson junction based qubits \cite{oliver20}, or of unknown structural defects in superconducting circuits \cite{pop23}. A set of harmonic oscillators in form of superconducting cavities would provide an alternative realization of a Caldeira-Leggett environment. Here we focus on TLSs. Returning to the archetypal setup, where a central qubit couples to an ensemble of these TLSs, we observe the dynamics of this qubit when initially set to its excited state. We use the same model as above, but now with different distributions of energies and couplings of the TLSs. For the sake of clarity of the argument, all the TLSs are again set initially to their ground state, mimicking a zero temperature environment. As we have shown in another context \cite{jb22}, a broad distribution of energies of the TLSs secures exponential decay of the qubit population in time. This can be seen also analytically, for instance, by standard means resumming in all orders of perturbation assuming a large number of uniformly distributed TLS energies. The distribution of energies and couplings of the TLSs is an essential condition for absorbing the energy of the qubit to this bath without recoveries over any practical timescales. In this case, the qubit decays exponentially as 
\begin{equation} \label{expo}
	|\mathcal{C}_0(t)|^2\simeq e^{-\Gamma_0t}.
\end{equation}
Here $\Gamma_0=2\pi\nu_0\Lambda_0^2/N$ with $\nu_0$ the density of TLSs around $\Omega$, and $\Lambda_0^2=\sum_{i=1}^N \gamma_i^2/\hbar^2$.

In general, for any distribution of energies and couplings, we find that the qubit amplitude 
$\mathcal{C}_0(t)$ in the excited state is governed by the integro-differential equation
\begin{eqnarray}\label{rebath}
	&&\ddot{\mathcal{C}}_0(t)+\Lambda_0^2~\mathcal{C}_0(t)=\\&&-\frac{i}{\hbar^2}\sum_{k=1}^{N}\gamma_{k}^2(\Omega-\omega_k)\int_{0}^{t}dt' e^{i(\Omega-\omega_k)(t-t')}\mathcal{C}_0(t').\nonumber
\end{eqnarray}
We see immediately that for the case where all the TLSs have the same energy as the qubit, $\omega_k\equiv\Omega$ for all $k$, the qubit does not decay, even when the couplings $\gamma_i$ are fully random, but it oscillates with population $|\mathcal{C}_0(t)|^2=\cos^2(\Lambda_0t)$, i.e. the Poincare recovery time is $\pi/\Lambda_0$. 

We can generalize the conclusion above for a bath where $\omega_k=(1-r)\Omega$ for arbitrary positive $r$, meaning detuned equal-energy TLSs in the environment. In this case, Eq. \eqref{rebath} leads to 
$\ddot D(t) -ir\Omega \dot D(t)+\Lambda_0^2 D(t)=0$,
where $D(t) = \dot{\mathcal{C}}_0(t)$. ${\mathcal{C}}_0(t)$ satisfies the initial conditions ${\mathcal{C}}_0(0)=1$, $\dot{\mathcal{C}}_0(0)=0$ and $\ddot{\mathcal{C}}_0(0)=-\Lambda_0^2$. We then have the oscillatory solution
\begin{equation} \label{eqe}
|\mathcal C_0(t)|^2=1-\frac{\Lambda_0^2}{\Lambda_0^2+(r\Omega/2)^2}\sin^2(\sqrt{\Lambda_0^2+(r\Omega/2)^2}\,t).
\end{equation}
\begin{figure}
	\centering
	\includegraphics [width=\columnwidth] {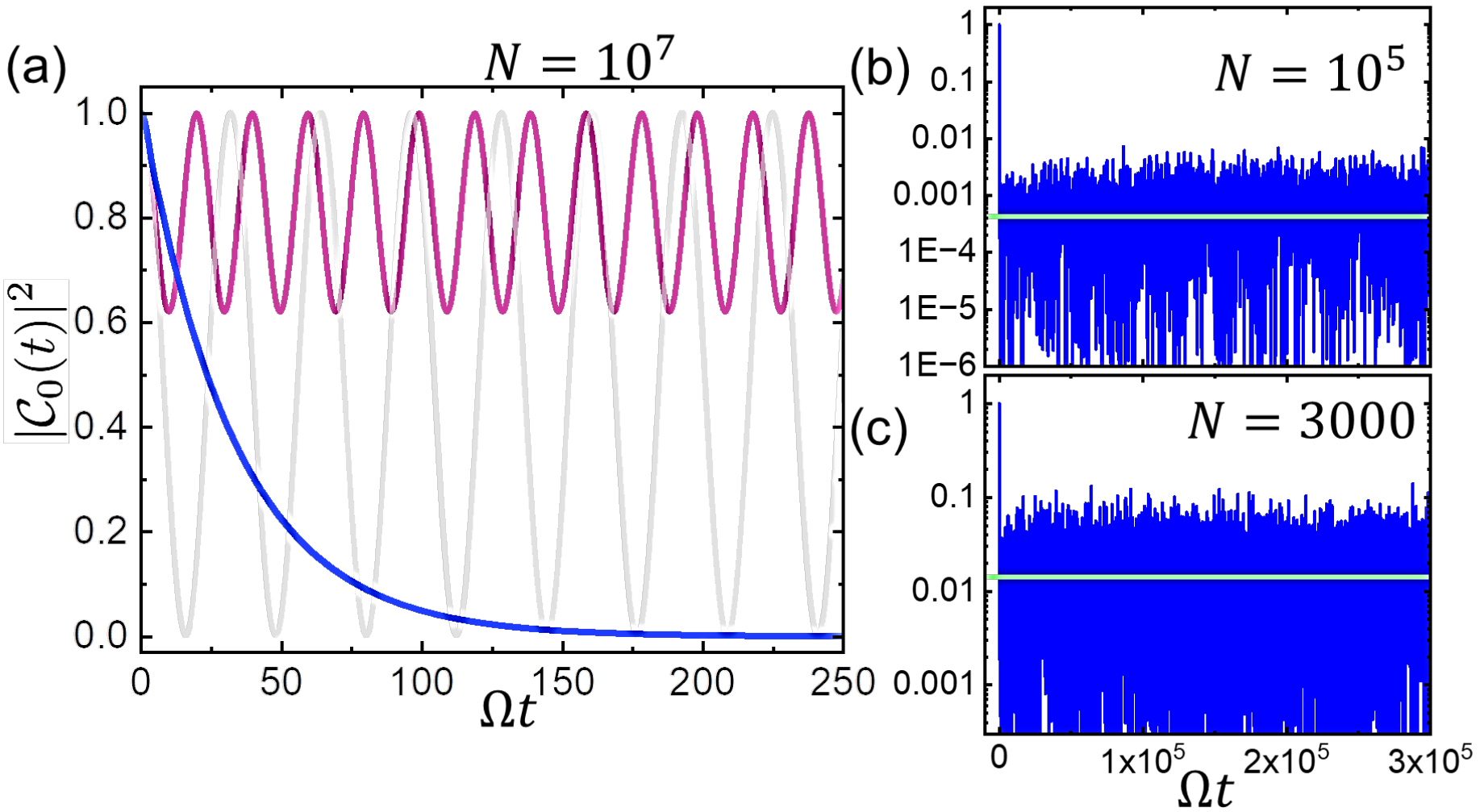}
	\caption{A qubit coupled to a reservoir of $N=10^7$ two-level systems in (a). The central qubit is coupled to each TLS via coupling constants $\gamma_i$ that have a uniform distribution between 0 and its maximum level, corresponding to the overall relaxation rate $\Gamma_0=0.03$. The dark blue line corresponds to the evolution of $p_e\equiv |\mathcal C_0|^2$ in the environment of TLSs with uniform distribution of energies in the range $0 < \omega_i <2\Omega$ leading to nearly exponential decay. The oscillatory qubit populations of the other curves correspond to uniform environments with $\omega_i = (1-r)\Omega$ for all $i$, with $r=0,0.25$ for grey and red lines, respectively. These dynamics follow that given by Eq. \eqref{eqe} quantitatively. (b) and (c) show the population in a similar distributed bath of $N=10^5$ and $N=3000$ TLSs, respectively,  over a time period of $\Omega t = 3\times 10^5$. The horizontal lines are the analytical long time predictions given in the text.
		\label{fig3}}
\end{figure}

Figure \ref{fig3} (a) shows the numerically calculated results of $p_e(t)$ for $N=10^7$ TLSs and for different choices of parameters following closely the analytical results given above. For a uniform distribution of TLS energies in the range $[0,2\hbar\Omega]$ the decay is exponential as described above, whereas for TLSs with identical energies there are periodic revivals, in quantitative agreement with the analytic result. These results serve as a warning sign for models where bath spins are assumed to have equal energies. In Fig. \ref{fig3} (b) and (c) we monitor numerically the long time behavior of $p_e(t)$ under the same conditions as in the main frame, but with $N=10^5$ and $N=3000$ TLSs with distributed energies and couplings. We see that there are no revivals over this long period of time in both cases, and the long time population follows closely the prediction $p_e(t\rightarrow \infty) =4\Omega/(N\pi\Gamma_0)$ indicated by the horizontal lines \cite{lindbladissue23}.

Two possible realizations of such reactive baths can be immediately envisioned. The one that corresponds to our analysis here is that of a qubit coupled to TLS environment with variable energies: with modern qubits as TLSs the couplings and energies can be varied almost arbitrarily \cite{oliver20}. One can envison to couple hundreds, perhaps even thousands of such artificial TLSs to a qubit. A simpler choice could be an ensemble of superconducting resonators with the same idea: here the tunability is more limited and instead of TLSs, these resonators work as harmonic oscillators.

In summary, it is possible to form a thermal bath on a chip avoiding recurrences \cite{bocchieri57} over any practical time scale in the spirit of Caldeira and Leggett \cite{caldeira83} using just reactive elements. However, a one-dimensional array of Josephson junctions or alternatively a transmission line exhibits periodic recoveries on nanosecond time scales in practical physical circuits for two reasons: first, the energy distribution is not dense and, equally importantly the coupling is not random but essentially equal ($\propto \sqrt{i}$) to each state $i$. Such an environment is thus a heat bath only if it has significant intrinsic dissipation, valid typically for $N > 10^5$ \cite{rastelli18,masluk12}, or if it is terminated by a resistive element \cite{cleland20}; in this case the termination itself is the bath. A way around to achieve a true bath is to form a network of harmonic oscillators or TLSs with distributed parameters and couple it to the quantum system.

We thank Diego Subero, Charles Marcus, Andrew Cleland, Youpeng Zhong, Vladimir Manucharian, Alfredo Levy Yeyati, Mikko M\"ott\"onen, Arman Alizadeh and Paolo Muratore-Ginanneschi for useful discussions. This work was supported by Research Council of Finland Grant No. 312057 (Centre of Excellence program) and Grant No. 349601 (THEPOW).


\begin{thebibliography}{99}
\bibitem{note} These frequencies practically coincide with those obtained by exact diagonalization of an array of arbitrary length \cite{rastelli18} and those from numerical results for $n \ll N$, which is the usual regime.
\end{thebibliography}
\end{document}